\documentclass[12pt]{article}

\newcommand{\ket}[1]{|#1\rangle}

\def\fract_HP#1/#2{\leavevmode
  \kern.1em \raise .5ex \hbox{\the\scriptfont0 #1}%
  \kern-.1em $/$%
  \kern-.15em \lower .25ex \hbox{\the\scriptfont0 #2}%
}%

\begin{document}
\title{\bf Is nature OO?}

\author{Guy Barrand\thanks{barrand@lal.in2p3.fr}\\LAL, Univ Paris-Sud, IN2P3/CNRS, Orsay, France}
\date {July 2011}
\maketitle

\begin{abstract}
  What exists "out there"? What does "doing physics" mean? What are the
 axiomatic ideas for microphysics? What is a particle? What is an apparatus
 made of? We show that Quantum Mechanics textbooks cannot truly answer
 this kind of question whereas they should. By adopting a pure "hitological"
 point of view for microphysics, we introduce the Hit in Apparatuses
 Theory (HAT) and the Vacuum of Apparatuses (VA) that restore, through
 Object Orientation (OO), an intuitive
 ontology to deal with this kind of physics. Through a review of what
 it means to "observe" and what relativism means in Special and
 General Relativities (SR and GR), we address the problem of finding
 common maths for GR and QM. Finally, with our new HAT, we address the
 measurement problem in QM and propose two possible approaches.

\end{abstract}

Keywords: microphysics, apparatus, hit, detector, source, hitology, vacuum, observer, relativism, object orientation, OO, HAT, VA.

PACS numbers: 03.65.Ta, 03.65.Ca, 03.65.Ud, 03.30.+p, 04.20.Cv

\section {What is "doing physics"?}
  Physics is made of three components: ideas about nature, mathematics,
 experiments. A good theory should have all three. First of all,
 clear ideas about something in nature, something "out there". Second, a clear
 mapping of these ideas with mathematical symbols. Third, a good matching
 of experimental results or observations (the data) with what is derived from
 the maths. We have the deep conviction that if we do not have all three,
 especially the right ideas, we are not doing good physics or even
 physics at all! This could be represented with the semantic equation
\begin{displaymath}
   doing\mbox{-}physics = (ideas,maths,experiments)
\end{displaymath}
\par
  Maths is the logical manipulation of propositions made with symbols.
 One set of propositions, the axioms, is more fundamental
 than the others. The axioms are considered de facto as true and are, by
 construction and definition, out of the process of demonstration because
 demonstrations and theorems rely on the axioms. Regarding the ideas for
 physics it appears that we have a similar kind of process. Among all
 ideas some seem to be more fundamental than others, for example the ideas
 of space and time. We are going to qualify such ideas as "axiomatic".
 It is interesting to question what the axiomatic ideas are
 in today's physics. For example, we are going to see that the
 idea of "corpuscle", defined as a little object "out there" bearing
 properties of its own and "flying around", is far from being such an
 obvious axiomatic idea as it looks.
\par
  It is interesting to note that this kind of reductive process seems
 to apply in experimental physics too. Some experiments, such as the
 "two slits" one, reveal so sharply a peculiar feature of nature
 that they look axiomatic too!

\section {The Hilbert Formalism (HF) }
\subsection {The Schr\"odinger equation }
  Before the mid 1920s we had very good theories dealing
 with meso and macro scales "out there": classical mechanics, 
 Maxwell theory, General Relativity. In particular in
 these theories ideas did not pose problems. They were
 the ideas of solid body, space, time, field. These ideas were
 nicely symbolized by using differential calculus.
 Moreover, we had an impressive match with experiments. However,
 things collapsed around 1925 with micro-scale physics.
 One key experimental fact was the observation of the spectrum
 of emitted light from an illuminated hydrogen gas that appeared
 to be\ldots discrete! A key theoretical fact was the
 discovery by Schr\"odinger of an equation able to recover
 this spectrum
\begin{displaymath}
  -\frac{\hbar^{2}}{2m}\Delta\psi+V\psi = E \psi
\end{displaymath}
 This equation (the "time independent" one) operates
 on a field $\psi$, has the symbol $E$ representing Energy as
 a parameter and a "potential field" $V$ as a constraint.
 A wonderful feature of this equation is that for some
 particular potentials it has solutions for $\psi$ only if $E$ is in
 a discrete spectrum, and it appears that in the case of a $1/r$ potential,
 this spectrum matches the observed one of the illuminated
 hydrogen gas. This is brillant (no pun intended). Especially knowing
 that the Maxwell theory, based on the idea of corpuscles flying
 around in a field mapped on space-time, was not able to recover
 the observed spectrum. The calculations can be found in many
 books. The author encountered them for the first time
 in \cite{ref-landau-3}. 
\par
  This magic equation seems to solve the question of microphysics, but 
 a "little detail" prevents that: we have no clear idea to attach
 to\ldots $\psi$!

\subsection {Definition of the HF}
  The Schr\"odinger equation leads to new maths that we are going to
 name here the Hilbert Formalism (HF in short). It is the
 maths used in today's microphysics. The HF is based on Hilbert vector space
 using complex numbers, equipped with an inner product to get real numbers
 from vectors, operators acting on a vector, a whole corpus of logic
 to analyse the spectrum of operators, a way to decompose a
 vector (a $\psi$) into the basis of an operator, etc\ldots In this text
 we are not going to question this formalism but more to question the
 ideas of the physics attached to it and, for example, the idea attached
 to the $\psi$ of the Schr\"odinger equation which is itself part of the HF.
 
\subsection {So what is the idea for $\psi$? Probability of what?}
  What is $\psi$ for? A textbook answer is that $\psi$ is used
 to calculate probabilities. Fine, but the probability of what?
 It would be natural to say that the axiomatic ideas for microphysics are
 space, time and corpuscles able to "fly around" and that $\psi$ is used
 to calculate the probability that a corpuscle "be" in a
 given position, with the randomness having
 its origin in some unknown effect to be discovered or, why not, being
 an axiomatic idea of randomness.
\par
  In general, experts, teachers and textbooks discard this view, this
 interpretation, by saying that $\psi$ is used to calculate the probability
 that the corpuscle "be found" at a given position. This is a subtle
 difference, but an essential one. In particular this "be found"
 assumes de facto an apparatus logic in the foundations, in the
 axiomatic ideas. This "be found", and the fundamental change of semantic 
 attached to it, has its origin in a piece of maths in the HF called the
 "Heisenberg inequality" and an interpretation
 of this inequality called the "Heisenberg Uncertainty Principle" (HUP).
 We are definitely not going to argue about this "principle"
 here (perhaps in another text when equipped with our new HAT),
 but a key feature of the HUP is to say
 that we can no longer attach both the properties of position
 and speed (momentum in fact) to a corpuscle. With the consequence
 that the idea of trajectory no longer makes sense in microphysics, 
 and so the idea of "flying around out there" makes no sense either.
\par
 In order to keep the connection to nature, textbooks fall back
 on an "axiomatic measurement logic" by attaching position or momentum to
 the result of a measurement process on a "tiny something" that we are going
 to name "particle" from now on rather than "corpuscle". We reserve the
 term "corpuscle" to a "tiny object" having a trajectory as in classical
 mechanics or Maxwell theory. We may write the definitions
\begin{displaymath}
   corpuscle \stackrel{def}{=} tiny\ object\ with\ trajectory
\end{displaymath}
\begin{displaymath}
   particle \stackrel{def}{=} tiny\ object\ without\ trajectory
\end{displaymath}
\par
 A key point is that a "particle be found" assumes anyway that there
 are "tiny objects out there" beside the axiomatic apparatus
 needed to attach $\psi$ to the idea of "be found".
 It is here that we start to have problem of interpretation because
 if $\psi$ is dedicated to a "be found", there remains nothing in the
 formalism to symbolise a "tiny something"! In particular, since there
 are no more trajectories, there are no $X,Y,Z(t)$ symbols to represent a
 "tiny object out there". So with these ideas attached
 to the Hilbert Formalism, we are driven to a strange interpretation in which
 some axiomatic ideas ("tiny object" or "particle") have no direct
 mapping symbols in the formalism. Weird.
\par
 Moreover, the symbols of space and time, t and $\stackrel{\rightarrow}{x}$,
 appear both in $\psi(t,\stackrel{\rightarrow}{x})$ and in the partial
 derivaties of the Schr\"odinger equation. A three dimensional
 uniform and constant metric is also hidden in the Laplacian $\Delta$ of
 the equation (through a $\delta^{ij}$). So we must conclude
 that a "found position" has to be understood from an apparatus immersed in
 a Euclidean space-time. This induces a more acute problem of
 interpretation since we have to speak of a "tiny object out there"
 not represented in the formalism, having no trajectory of its own, and
 that cannot be said to be "here", but is nevertheless
 in a "here" when measured because of the Euclidean "here" defined by the
 apparatus! Highly weird.
\par
 To labour the point we could mention the "spin". If we put an illuminated
 hydrogen gas in a magnet we observe a change in the emitted spectrum.
 It appears that the HF has the spinor operator piece of maths that brings
 the necessary degrees of freedom to model the modified spectrum.
 So we have a good match of the maths with an experimental result - fine!
 However, things go wrong when looking at the ideas presented in textbooks
 to justify the usage of the spinor maths.
 Spinor is a mathematical object related to rotation in a Euclidean
 three dimensional space and textbooks attach
 a "spin" property to their "particle" (spin $1/2$ of the electron).
 How can we attach to the "tiny object" a property
 related to rotation in the Euclidean space of the apparatus
 which is the only space at hand in this interpretation? How can we attach
 the idea of meso scale rotation to something that cannot be said to be "here"?
 How can we attach a geometric idea to a "tiny object" that cannot be
 said to "be" in space? Most textbooks evade the issue
 by saying that spin is a "pure Quantum Mechanics effect" which obviously
 explains\ldots nothing! Worse, before reaching this conclusion
 some dare to use the analogy of the classical spinning top
 in order to give an "intuitive feeling" to what the "spin" of a
 "particle" is. A disaster! (A classical spinning top is an object having
 a spatial structure that rotates "out there").
\par
  If there is no $XYZ(t)$ in the formalism, and no more symbols
 to represent the "particle" directly, it would be much more consistent to say
 that there are no particles at all and then say that $\psi$ is used
 to calculate the probability that a cell
 at a given position in a measuring apparutus has to fire. This is much more
 convincing since the HF has symbols to describe a measuring device
 (the operators) and also has symbols to describe a "source apparatus"
 (the vectors). "Source" here is interpreted as the source of the firing
 events in the sense that, if the source apparatus is not there, then
 we never observe the firing of cells.
\par
  It is this "no particle" idea that we would like to develop in this text
 to see if we can have a more consistent approach to microphysics. More
 consistent at the level of the maths and experiments, but also of the ideas.

\subsection {What is Quantum Mechanics?}
  A textbook about microphysics that uses the HF is said
 to be about "Quantum Mechanics" (QM in short) \cite{ref-landau-3},
 \cite{ref-messiah}, \cite{ref-cohen}. As just discussed,
 a particle vocabulary is still heavily used in these QM books.
 The words atom, electron, particle, etc\ldots often appear
 in the introduction as if taken for granted, without any kind of definition.
 To help the argumentation of this text, we are going to rely on
 this usage of a particle vocabulary to define "Quantum Mechanics".
 In fact few books come with a definition of what QM is.
 For us, QM is the area of science dealing with microphysics
 based on the maths of the HF but still associated with a particle
 vocabulary.
\par
  Having the conviction that this undue vocabulary is
 the source of the intuitive discomfort that many people
 have with "QM" (including the author), we are going to see
 if it is possible to reread the HF by avoiding this vocabulary
 and then restore a clear understanding of the science
 of microphysics.

\section {HAT: Hit in Apparatuses Theory}
\subsection {HAT, detector and hit-source definitions}
  HAT, for Hit in Apparatuses Theory, is defined as an interpretation
 of the HF formalism based solely on the two axiomatic ideas of apparatuses
 and hits appearing in them.
\par
  Experimental microphysics shows that we can classify apparatuses
 in two categories, the "detectors" and the "hit-sources".
 We define a "detector" as an apparatus in which hits appear. For the
 moment the hits are zero dimensional (punctual) events appearing in
 the detectors. Some apparatuses are such that, if they are not present,
 no hits appear in a detector apparatus. We define
 a hit-source apparatus as such an apparatus. In a complex
 detector (some made of many devices) hits often appear in a
 pattern that characterizes the hit-source (for example an
 "electromagnetic shower" hit pattern).
\par
  It is important, in fact fundamental, to note that the definitions above
 do not use the words corpuscle and particle at all.
 We consider that these definitions are good foundations because
 they do not pose a problem of realism for us. At first glance, HAT looks
 like some kind of "hitology", but we are going to see that
 it is more than that.

\subsection {A no-go for corpuscles in microphysics?}
  To define apparatuses we could
 have said that they are made of an aggregate of corpuscles (as defined
 in the first paragraph), that a hit-source is a device that emits corpuscles
 and that a detector is a device that reacts by producing a hit
 when impacted by a corpuscle.
 "Corpuscle" would then have been a universal axiomatic idea. It is
 very natural to attempt to define apparatuses and the whole of microphysics
 in this way, but it appears that one part of this kind of
 theory\ldots does not work!
\par
  A nice reasoning of John Bell, tested
 in "Bohm-Aspect" kind of experiments, is said to rule out the idea
 that a hit is produced by a corpuscle emitted from the source. 
 For clarity of the overall argumentation, we must explain, with our words,
 the outline of a Bohm-Aspect type experiment and the Bell reasoning.

\subsubsection {Bohm-Aspect setup}
  We can imagine an experimental setup composed of three apparatuses,
 a hit-source and two detectors placed remotely aside
 the hit-source by having the three devices aligned along the same
 axis (named z here). Each detector is such that it defines
 an oriented axis in a plan perpendicular to z so that
 "+hit" can appear in the forward direction of this axis and "-hit"
 can appear in the backward direction of the axis.
\par
  For a particular kind of hit-source found in nature,
 we can observe time coincident pairs of hits in both detectors.
 The pairs are of four kinds: (+,+) (+,-), (-,+), (-,-).
 We can count the number of pairs with the same sign
\begin{displaymath}
      N_{same}     = N_{++}+N_{--}
\end{displaymath}
 and with the opposite sign
\begin{displaymath}
      N_{opposite} = N_{+-}+N_{-+}
\end{displaymath}
 and then calculate the "correlation factor" C as
\begin{displaymath}
      C = (N_{same} - N_{opp})/(N_{same} + N_{opp})
\end{displaymath}
 One macroscopic parameter of the setup we can play with,
 is the relative angle $\theta$ of the axes of the two detectors. Other
 macroscopic parameters are the two distances
 of each detector to the hit-source, but strangely they do not appear in the
 reasoning. We can then do various acquisitions (runs)
 by getting C for various $\theta$ and draw an experimental
 curve $C_{exp}(\theta)$. As the shape of this curve is not
 relevant for the argumentation, we are not going to show it here. It helps
 to concentrate on the essentials.
\par
 It is important, in fact fundamental, to note that in the description of the
 experiment we have not used the words corpuscle, particle and
 in particular "pair of photons".

\subsubsection {Bell reasoning}
  It is at this point that Bell's reasoning comes into play.
 Bell claims that a large set of theories describing the experiment, and
 in particular the ones based on corpuscles, must match some conditions,
 the Bell conditions (BC in short), and that when these conditions are met,
 then the $C_{theory\ under\ BC}(\theta)$ has some constraints. The passage
 from the Bell conditions to the constraints over $C(\theta)$ is the
 Bell theorem. A striking result is that these constraints are such that
 a $C_{theory\ under\ BC}(\theta)$, and then $C_{corpuscle\ theory}(\theta)$,
 cannot reproduce the $C_{exp}(\theta)$! 
\par
  In Bohm-Aspect-Bell (BAB in short), it is not the fact that there
 are coincident hits that poses a problem, but the fact that the
 amount of correlation for some $\theta$ cannot be explained by
 some theories, and in particular
 by the most intuitive theory that we can imagine at first, the one describing
 nature with corpuscles flying around and interacting locally with the
 detectors. Said simply, for some $\theta$s there is too much correlation
 for an intuitive corpuscle theory. With the BAB logic,
 it seems that we have a proof that this kind of theory cannot work and
 therefore that our intuition is baffled by experimental microphysics!
\par
  After the pioneer texts \cite{ref-bohm}, \cite{ref-aspect},
 \cite{ref-bell}, a lot was written to criticize\ldots everything!
 In general criticisms are of two kinds. First, criticisms around
 how the experiments are done. Second, criticisms around the fact that
 the Bell conditions cannot put aside all the corpuscle based theories.
 In this text we are going to assume that "BAB is granted"\footnote{
 It is not so clear whether the Bell conditions cover the case
 of a theory based on corpuscles flying around in a space-time
 which is not "gently flat" at micro scale, a space-time having some dynamics
 of its own that could be viewed as the origin of the "too much correlation".
 We assume in this text that this kind of theory is ruled out too.}
 , and in particular that the experimental data and the $C_{exp}(\theta)$
 curve are granted.

\subsubsection {The HF at work and the HAT point of view}
  This loss of the idea of corpuscle looks like the end of
 "doing microphysics", but the situation is partially rescued because\ldots
 it is possible to model
 this experiment with the Hilbert Formalism! If the hit-source is modeled
 by a vector of the HF and the detectors are modeled by operators of the HF,
 the formalism makes it possible to calculate a $C_{HF}(\theta)$ that matches
 the $C_{exp}(\theta)$! Since the setup was presented
 by using a pure HAT terminology, and we have defined HAT has being
 associated to the HF, then we can write 
\begin{displaymath}
      C_{HAT}(\theta) = C_{HF}(\theta) = C_{exp}(\theta) 
\end{displaymath}
 and so we have restored clear ideas mapped to neat maths that recovers
 the data : "doing microphysics" is back for this experiment!
 Moreover, it is back in a way that reinforces
 a pure HAT point of view since the Bell reasoning is said to eliminate,
 in this case, the word "corpuscle".

\subsubsection {The QM point of view}
 As QM is also attached to the HF, we have
\begin{displaymath}
      C_{QM}(\theta) = C_{HF}(\theta) = C_{exp}(\theta) 
\end{displaymath}
 but what is striking is that in general the experimental setup is presented
 by using the words "pair of photons" to qualify the "source". As QM defenders
 also accept the "loss of corpuscles"
 coming from the Bell reasoning, we are driven into a strange microphysics
 in which on the one hand the "good old corpuscles" are said to be ruled out,
 but on the other hand the word "photon" is nevertheless used to describe
 the setup! Weird\ldots again!
\par
  To qualify this strange "pair of photons source"
 that can produce, for some $\theta$, an amount of correlation 
 not reproducible by a corpuscle theory, the word "entanglement"
 was introduced (the source is often presented
 as a source of "pairs of entangled photons"). This new word obviously 
 clarifies nothing, since we have no clear idea of the nature of the entity
 being qualified! For us, the questioning around this kind of experiment
 is not to qualify a source of "photons", but
 to know if it still makes sense to use the word "photon" at all!
 QM defenders should first speak about an experimental setup with
 a "hit-source" being able to produce particular coincident hits,
 and then ask the question: does a photon entity make sense to explain them?
\par
  It is worth noting that the HAT point of view transforms
 an uncomfortable feeling of weirdness coming from QM, to a healthy feeling
 of awe. The awe at finding in nature hit-sources
 able to produce such $C_{exp}(\theta)$ not explainable by a corpuscle theory!

\subsubsection {A remark}
 It is also interesting to note that the BAB argumentation does not destroy
 the idea of apparatuses being, or not being, an aggregate of corpuscles!
 Strictly speaking BAB does not address this problem, it destroys the idea of
 corpuscles only for the "in between" apparatuses. Consequently, we
 start to realize that the nature of apparatuses is going to be a central
 question. This point is going to be explored later.

\subsection {We are Object Oriented!} 
  For us, being unable to decide on which foot to dance
 with the word "particle" in QM is what induces the huge discomfort
 that we have with this interpretation of the HF for microphysics.
 The discomfort arises because the idea of property-bearing objects is
 something deeply rooted in the way we think: we, as human beings, are\ldots
 object oriented! We are "OO", and a theory about nature unable to
 pinpoint its own objects cannot be a good theory for us.
 We claim that HAT is better than QM, because HAT clears the decks
 concerning the word "particle".
\par
  And what if nature were not OO? If that were the case, we would be unable
 to find the right ideas for the "out there", which would mean a true end
 to "doing physics" as defined above, but it seems that we still have
 some cards to play, so let us continue\ldots

\subsection {The two slits experiment}
  The "two slits" is a canonical experiment used in QM textbooks
 to justify the HF. This justification comes from the fact that the
 HF contains a vector addition which represents very well
 what is observed. In general the two-slits
 is also presented as the canonical experiment
 showing that "microphysics is weird", and this because there is
 no way to answer the canonical question "through which slit does the particle pass?".
 As we are going to see, a HAT point view naturally removes any kind
 of weirdness here.
\par
  As for the Bohm-Aspect setup, it is important,
 in fact fundamental, to be careful about the words used to present
 the experiment. The setup is made of a hit-source apparatus pointing
 in a direction z, a farther cache perpendicular to z with two
 parallel slits (A and B) and a farther plane detector also
 perpendicular to z. The four experimental situations 
\begin{enumerate}\addtolength{\itemsep}{-0.5\baselineskip}
   \item slit A opened, B closed
   \item slit A closed, B opened
   \item slit A opened, B opened
   \item slit A closed, B closed
\end{enumerate}
 could be modeled with a $\sigma$=1,2,3,4 macroscopic parameter.
 In the reasoning, this parameter is an equivalent of the $\theta$ macroscopic
 parameter of the Bohm-Aspect setup. For the first three cases, according
 to $\sigma$ (and then for different runs labeled by $\sigma$), we observe
 three distributions of hits: $D_{exp}(\sigma=1)$, $D_{exp}(\sigma=2)$,
 $D_{exp}(\sigma=3)$. An interesting fact is that in the case $\sigma$=3, the
 distribution of hits has an "interference" pattern, whilst
 each distribution $\sigma$=1,2 does not (both are circular). So we have
\begin{displaymath}
   D_{exp}(\sigma=3) \neq D_{exp}(\sigma=1)+D_{exp}(\sigma=2)
\end{displaymath}
 What is nice is that we can model these three situations quite easily
 with the HF by associating a $\psi$ for each $\sigma$. A mathematical
 curiosity is that, apart from a normalization factor, we have
\begin{displaymath}
   \psi(\sigma=3) = \psi(\sigma=1) + \psi(\sigma=2)
\end{displaymath}
 and that $\psi(\sigma=3)$ recovers the interference pattern.
 So far so good, and we could have stopped the presentation of the two-slits
 here since the three ingredients of "doing physics" are here. Clear ideas
 (apparatuses, hits), good maths (the $\psi$s and the capability to add them)
 and a very good matching with experiment (in particular the recovery
 of an interference pattern).
\par
  An important and fundamental fact is that until now we have not
 used the word "particle", nor the word "corpuscle", and that so far
 the above two-slits presentation is clear. Now if attempting to model
 this experiment with a corpuscle theory,
 we fall on a serious problem because a "standalone corpuscle flying around"
 theory would lead to
\begin{displaymath}
   D_{corpuscle\ theory}(\sigma=3) = 
    D_{corpuscle\ theory}(\sigma=1) + D_{corpuscle\ theory}(\sigma=2)
\end{displaymath}
 which is not what is observed. In particular, a corpuscle theory would not
 lead to an interference pattern. So, as in BAB, we are driven to the
 conclusion that the idea of corpuscles is ruled out in this microphysics
 experiment. In fact, we could have used the two-slits as a corpuscle
 no-go argumentation instead of BAB, but BAB is more interesting
 since it eliminates more theories. It should be noted that it is not so
 much the fact that there is an interference pattern in $\sigma$=3
 which is important as the mere fact that $D_{exp}(\sigma=3)$ is not the same
 as $D_{exp}(\sigma=1)+D_{exp}(\sigma=2)$. This non equality alone is
 sufficient to conclude.
\par
  In QM textbooks or lectures, it is highly instructive to study the section
 on how the two-slits experiment is presented. Most of the time, it is
 presented in the first lecture by using the words "particle" or "electron"
 as if taken for granted. The "source" apparatus is presented de facto
 as a source\ldots of particles, themselves often presented as
 corpuscles (sometime even drawn on the blackboard !). This is wrong, and
 because of this usage of the wrong vocabulary so early, the poor student
 cannot avoid catching an intuitive discomfort right from the first lecture,
 a discomfort that leads in general to strong nausea by the end of the term!
 The two-slits is presented so early more to sell the HF
 than anything else, in particular the linearity of the algebra,
 the fact that $\psi(\sigma=1) + \psi(\sigma=2)$ has a physical meaning
\footnote{Students that feel comfortable are in general more mathematicians than
 physicists and do not run away (fast) because the HF, with its linear
 algebra, is a nice piece of maths to play with. To be fair, we agree on
 that.}. 
 But the point with the two-slits experiment is not in the maths!
 It is in the fact that this experiment is a canonical one to question
 the usage of the words "particle" and "corpuscle" in microphysics.
\par
  We also see that a pure HAT, a pure hitological point of view,
 clarifies the question "through which slit does the particle pass?".
 HAT leads immediatly to the conclusion that
 this question is not answerable because it is\ldots ill defined!
 It is ill defined because the word "particle" is ill defined in this context.
 The HAT point of view also transforms the sentence "microphysics is weird"
 to "microphysics is awesome". It is awesome because we can
 find in nature, "out there", hit-sources able
 to produce an interference hit pattern and we can model the experiment
 by using a nice linear algebra. Truly marvellous!
\par
  The HAT point of view makes it possible to raise an interesting question:
 what about the case "A closed and B closed" ($\sigma$=4)? Our hitological
 point of view does not rule out the possibility of actually observing
 hits in the detector! Before saying that this is impossible, we must
 remember\ldots the tunnel effect.

\subsection {MachZender (and delayed choice) experiments}
  We could also have mentioned the MachZender "two arms interferometer"
 kind of experiment that would have drawn the same conclusions
 as for the Bohm-Aspect and two-slits ones.
 The macroscopic parameter to play with would have been the difference
 of length ($\delta$) between the two arms. In such an experiment, various
 runs according to this macroscopic parameter would have induced
 some $D_{exp}(\delta)$ experimental curve not reproducible with a
 theory of corpuscles flying around, but reproducible with the HF.
\par
  Here too, we would have concluded that the question
 "through which arm did the photon pass?" is ill defined
 and then unanswerable because the word "photon" is improper in this context.
\par
  A QM point of view would have shown the same defects as
 for the two-slits : a too early and undue usage of the word "photon" and a
 focalization on the maths. The conclusion would had been the same : the point
 is missed.

\subsection {Corpuscles? at what cost?}
  It must be mentioned that some models exist which attempt to model
 the two-slits or the Bohm-Aspect results by keeping corpuscles. The Bohm model
 is one of them (there are $X,Y,Z(t)$ with Bohm). Nevertheless all of them,
 at some point, have to introduce some weird ideas such as action
 at distance. Such ideas are definitely counter-intuitive and at some
 point not really OO. For example, action
 at distance induces that a corpuscle does not really bear properties
 in a standalone way since its behavior depends also on "the rest"
 (it is the "Wholeness" idea of Bohm-Hiley \cite{ref-bohm}).

\section {A pure "hitology"? No}
  Does the HAT interpretation, because it is an interpretation,
 induce that microphysics is reduced to a pure hitology? That
 is to say that nature is made of hit-source and
 detector apparatuses, all modeled with the Hilbert formalism
 for which the only goal is to calculate probability
 distributions of hits? In fact no, one particular
 set of experiments, the "decay" ones, induces that we have to consider
 that there is an extra entity in the whole story, the "in between"
 apparatuses, which appears to be an active physical entity.
\par
  To describe a "decay experiment", we first have to label a hit-source.
 It appears that in nature apparatuses exist, or can be built, that
 produce different kinds of hit pattern when the detector is placed very close
 to the hit-source or even without any space between the two.
 These different patterns make it possible to classify the hit-sources:
 electron-hit-source, photon-hit-source, muon-hit-source, etc\ldots
 Note that here the words electron, photon, muon are introduced with a
 definition (through a physical procedure). Few books
 in microphysics do that. We are going to name this definition, which is based
 on experimental facts produced with a particular apparatus setup,
 a "definition setup". We insist that these words
 are not, definitely not, introduced by describing some corpuscle
 entity "out there". It is interesting to note that to define the words
 electron, photon, muon, etc\ldots it was necessary to introduce pairs
 of apparatuses, pairs of $(hit\mbox{-}source,detector)$. A hit-source or
 a detector apparatus alone cannot do the job. This will be discussed
 again later.
\par 
  Armed with this definition and classification of hit-sources,
 we can observe that it is possible to find (or build) in nature
 the following setup. A hit-source can produce a first kind of hit pattern
 in its associated detector placed close to it, but can produce a different
 kind of pattern when the detector\footnote{For simplicity we assumed a same detector for the two patterns} is placed farther from it!
 Moreover, the pattern is not only changed by some geometrical
 factor (for example a different size of "electromagnetic shower"
 hit pattern) but can also be transmuted to a hit pattern
 which is associated to another kind of hit-source! A typical case is
 with a muon-hit-source and an electron-hit-source. If a detector is placed
 close to a muon-hit-source, we observe a muon-hit pattern but if the detector
 is placed some meters farther away we no longer observe a muon-hit
 pattern but an electron-hit pattern!
\par
 So what? A textbook explanation for this transmutation
 is to say that a corpuscle (for example a muon) is flying around
 and that it transforms itself in mid-flight into something else
 (an electron and two neutrino corpuscles in case of the muon).
 Now if, because of BAB, we cannot keep the corpuscle idea, then
 we are compelled to conclude that beside the source and detector
 apparatuses there is, in between them, an extra entity that plays
 the active role of transforming the observed hit patterns, and does this
 according to the relative position of the apparatuses at our human scale.
 To further analyse this "in between entity",
 we have to find a name for it, and we have the right to name it
 because we have found experimental facts that reveal the existence of
 this entity. We have to take care in choosing the name. In particular
 the name must reflect the fact that we deal first with apparatuses, that
 apparatuses are axiomatic ideas. The best
 name that we have found so far is the "vacuum of apparatuses",
 the VA in short. ("In Between Apparatuses", or IBA, could
 be a good name too).

\subsection {The Vacuum of Apparatuses, the VA}
  This name has the huge advantage of using two words that bear
 clear sense for us. In particular this is much better than attempting
 to name the in-between entity by "quantum vacuum", a name that uses the
 word "quantum" which has been so ill defined since 1925! We claim high
 and loud that having identified the in-between entity as an active one
 by using the concepts of apparatuses and hits, and having been able
 to name it with clear words is a huge conceptual
 step in the story of seeking the right ontology, the right objects, for
 microphysics. So to the question: is microphysics only a hitology?
 We can now answer no, it is not. Microphysics must
 be viewed, because of decay-like experiments, as the study of apparatuses
 and of the outsider VA. We see also that "decay" is a highly misleading
 word since relying on a particle idea. In the following text we are going
 to use "hit-transmutation" experiments instead.
\par
  The VA makes it possible to restore object orientation in microphysics.
 The VA is "something out there" that has properties of its own, and one of
 these properties is to transmute hit patterns. We also see that the
 VA is related to space defined through the relative position of apparatuses.
 Nevertheless, because of the hit pattern transmutation phenomenon, this space
 cannot be reduced only to geometry. It is more than geometry, and we can
 already conclude, without any maths, that this phenomenon clearly rules
 out any theory, as the two relativities, that attempts to model space
 (space-time in fact) between apparatuses by pure geometry alone.

\subsection {What is the maths for the VA?}
  In the HF, between the $\psi$ of the hit-source and the operator for the
 detector, there is an extra entity called the Hamiltonian
 operator. By using the time-dependent Schr\"odinger equation, the
 Hamiltonian operator transforms (evolves) the $\psi$.
 By doing a spectral decomposition of the evolved-$\psi$
 against the local $\psi$s of the detector attached to each outcome (cell),
 we can calculate the observed probability distribution of the firing
 events (the hits). The Hamiltonian is clearly describing
 something in between the hit-source and the detector apparatuses.
 Therefore it is natural to attach it, in our hitology interpretation,
 to our VA. So each of the essential components of the HF now receives its
 interpretation. We claim that these interpretations are based on
 better grounded ideas than the ideas found in QM (QM as defined above).

\subsection {The maths for the VA of a hit-transmutation experiment}
  A simple Hamiltonian, such as the one describing the "harmonic oscillator"
 in QM, cannot model the transmutation of hit patterns. We need more
 sophisticated mathematics for that, and it appears that
 this maths already exists! It is nothing more than the maths
 of a "Quantum Field Theory" (QFT) and in particular the maths of QED
 for the microphysics of electric-charge-hit-sources. However, the QFTs suffer
 the same problems as QM at the level of the ideas that refrain a clear
 understanding of them. Mainly the QFTs still make heavy use of a particle
 vocabulary.
\par
  This vocabulary is visually reinforced by the intensive usage
 of the Feynman diagrams.
 A Feynman diagram is perhaps a nice trick for doing a perturbative
 calculation, but it is a huge intuitive and ontological trap from the very
 moment that the branches are attached to the idea of particle and that
 the word "particle" is suspicious.
 Moreover, the QFTs introduce new words such as "virtual", "quantum field"
 and the winner "quantum vacuum", that lead straight to the trap.
 The word "virtual" qualifies a particle attached to a branch
 of a Feynman diagram, but it is definitely
 not clear whether the "virtual particle" is something "out there"
 or not! If "quantum vacuum" is associated to "no particle", and
 that particle is suspicious, then quantum vacuum is suspicious too.
 The best that we can do here is to say that "quantum vacuum" is the name of
 the maths symbol $\ket{0}$ found in the maths of a QFT, that's all.
 No clear idea can be associated to these two words.
 The same for "quantum field"; here too the best solution is to say that it is
 the name of an operator in the maths of a QFT. About "quantum vacuum",
 if people attempt with these words to qualify the in-between apparatuses
 (as for the "in between" of the two plates of a Casimir setup), we claim
 that our VA terminology is superior since much better defined.
\par
 This being said, a strong point
 with QFTs, and especially QED, is that their maths is very impressive
 in giving the right probability distributions, and especially
 the ones of hit-transmutation experiments.
 So we are perhaps in a situation where we have found the right maths
 but not yet the right ideas for them. Now let us see if we can
 reread QFTs with our
 hitology ideas. If we look closely, the relationship of a QFT with
 experimental physics is established only through an input and an
 output "state". In general the word "state" is presented by using
 a particle terminology, for example
 an input or an output state with an electron and a positron in it
 with their own 4-momentum. In the formalism, it is modeled with symbols
 such as $\ket{e^-e^+}$.
 Now if BAB is right, we can no longer retain such an idea since
 the idea of a particle is no longer relevant. Instead, we have to
 rethink the symbols above as modeling some apparatus able to produce
 a hit pattern characterizing the association of an electron-hit-source and a
 positron-hit-source as defined previously through their
 definition-apparatus setup. Note that the hit-source apparatus could be
 something very complex. It could be a full accelerator setup! For example
 the LEP machine in the 1980s, or the LHC for the symbols $\ket{pp}$.

 The final state, which is modeled with the same kind of symbols,
 has to be conceived as something attached to a hit pattern in a whole
 detector such as the ALEPH detector during the LEP era or the ATLAS
 detector at the LHC. In the formalism, to pass from an input state to
 an output state, there are a lot of in between operations and symbols
 that appear. What is the ontologic status of these in-between maths symbols?
 We are going to associate all of them as a model of the VA, which is
 something that exists for us and is very well defined as an object
 for us to work with (so unlike a "quantum vacuum"). 
\par
 Is there some specific set of symbols that maps the VA? In fact yes,
 we already have that. In a QFT everything is encrypted in the "Lagrangian".
 It is from this entity that in-between
 manipulations are derived and that final probability distributions
 are calculated. So the Lagrangian can be seen as the piece of maths
 representing the VA. We must point out that for us the symbols for the VA
 is not $\ket{0}$, since $\ket{0}$ does not bear any transmutational property.
\par
  This being said, we have now a better understanding of the meaning
 of the maths of a QFT and what QED is about. Moreover, the complexity
 of the maths attached to the VA reinforces the idea that this entity is
 far from being a "gentle space-time continuum" as described by the
 two relativities (Special and General), and various other space-time
 oriented theories.
\par
 In general we remain amazed at the complexity of the
 maths dealing with microphysics (and representing the VA for us).
 This algebraic inflation, originating from the introduction of the "i"
 of complex numbers by Schr\"odinger in his time-dependent equation,
 culminates in the SUSY maths where we end up
 manipulating extra dimensions made of Grassmann numbers.
 (Do SUSY defenders really believe that there are extra dimensions made
 of non-commuting numbers "out there"?)

\subsection {What is high energy or particle physics?}
  We can now have a better understanding of the part of science called
 High Energy Physics (HEP) or\ldots particle physics! Experimental HEP
 is nothing more than the construction of detectors and
 accelerators seen as hit-source apparatuses, the classification of hit
 patterns, the classification of natural hit-sources and the study of
 the hit pattern transmutations.
\par
 The theoretical aspect of HEP consists of finding the right QFT
 with the right Lagrangian that encodes all possible
 hit pattern transmutations found so far and makes possible the calculation
 of the right probability distributions of hits in detectors.
\par
 There is a lot to be done and someone can spend a whole research career
 in HEP science!

\subsection {What is the "Standard Model", the "Higgs"?}
  The "Standard Model of particle physics" (!) can be defined now as the best
 Lagrangian discovered so far that encodes all known hit-sources and
 observed hit patterns.
\par
  An interesting point in the QFTs is that a "Lagrangian logic" of its
 own appears in them. If we take a Lagrangian, it may be deduced from
 another Lagrangian with less symbols, in particular by applying
 transformations justified by mathematical symmetry criteria.
 The Lagrangian of the Standard Model is such a "less symbols" Lagrangian.
 To model correctly all the "weak decay" hit patterns,
 the "reduction of symbols" procedure requires also the introduction
 of a "Higgs" term that can be related to a hit pattern of its own, but
 a hit pattern not yet seen in any experiment!
\par
  For us, finding a "Higgs hit pattern" is the whole point of
 "seeking the Higgs" at the LHC. In particular "seeking the Higgs" cannot be
 "looking for a new little thing flying around". The "Higgs" is going to be
 a new hit pattern never seen before in any experiment, a pattern that will
 guarantee the mathematical consistency of the "best Lagrangian discovered so far", a Lagrangian that should be interpreted as describing a\ldots vacuum
 of apparatuses!
\par
  Does the Higgs term explain or solve everything at the conceptual level?
 In fact no, far from it. The Higgs term does not solve the integration
 of gravity in microphysics and it does not address, as BAB does, more
 fundamental issues concerning our understanding of microphysics. 

\section {Restoring OO in microphysics}
  We have seen that we can restore object orientation, and then
 good intuition, when dealing with microphyics. A first step is
 to get rid of the words corpuscle, particle and probably wave
 since nothing in our apparatuses measures or detects waves. Restoring OO
 could be done by the drastic rethinking, revisiting, of all the vocabulary
 used so far when dealing with microphysics. If some words are to be retained
 (such as electron, photon, atom,\ldots) they must be carefully defined, or
 redefined, by using a set of axiomatic words (apparatus, hit, vacuum
 of apparatus) that make sense for us. We claim in this text
 that this is possible by rereading the Hilbert Formalism as a
 hitology completed with the VA entity.
\par
  After having helped to recover intuitive
 comfort in microphyics, we are going to see that this hitological point
 of view can help in one of the outstanding problems in today's physics;
 the problem of the unification of meso-macro-physics with microphysics.

\section {Gravity, General Relativity and the Grail of Unification}
  The maths of the QFTs is the best candidate we have for microphysics.
 This maths makes it possible to recover all hit probability distributions 
 observed so far, and this, sometimes, with astoundingly accurate precision.
\par
 Nevertheless, gravity still eludes QFTs. Here we use "gravity" as a word
 originating in our every day experience in the meso scale and, as such,
 which does not pose a problem to our intuition. The best
 maths we have for this phenomenon observed at meso and macro scale,
 is the maths of General Relativity (GR in short). GR models this
 phenomenon as a curvature property of a Riemann continuum
 in which physical quantities are mapped on tensors.
 The main idea of GR is that gravity can be explained as an effect
 of space-time which is seen as an entity having a dynamics
 of its own. Awesome! With GR, space-time truly becomes an object
 with properties. GR is OO and space-time is one of its objects.
\par
  For meso and macro scales, GR theory is a brilliant example of "doing physics"
 as defined in the first paragraph. First, we have clear and elegant
 ideas; there are bodies ($X^\mu(s)$ in the maths) and fields "out there"
 embedded in a space-time which is an object of its own.
 Second, we have good maths, such as tensors and Riemann geometry, with
 a nice mapping of ideas to maths symbols; in particular space-time is mapped
 to a metric tensor field. Third, we have a good match with experiments and
 observations in the meso-macro scale domains. Defenders of GR mention
 a match up to $10^{-14}$ precision for the period of pulsar $PSR\ 1913+16$
 \cite{ref-penrose}.
\par
  However, GR does not cover a good part of microphysics and in
 particular the hit pattern transmutations. GR is not a theory of
 microphysics. For example, it cannot explain the discrete spectrum of
 an illuminated hydrogen gas. For almost a century, physicists have been
 grappling with a difficult problem: we have a good set of maths for
 meso-macro scales
 and another set of good maths for microphysics but each has
 a logic of its own and we have not yet found some appealing common
 foundation maths to bring them under one common banner! Being able
 to do that is the challenge of the unification of GR and QM.
\par
  It is interesting to note that most unification attempts are
 done at the mathematical level where theoreticians attempt to bring under
 the same algebra the maths of the HF and the maths of GR.
 For us it is not so surprizing that these attempts at "unification by maths" 
 failed so far. It seems that we forget that we deal with physics,
 and that a part of physics is ideas about what is "out there".
 We don't quite see how we can unify at the level of the maths if we have
 not unified at the level of the ideas! If the problem resisted
 for so long it is probably because we have not yet put the finger
 on the right set of ideas that would lead to a common underpinning maths for
 micro-meso-macro scales. Manipulating maths symbols having no mapping
 to an idea about something in nature is not doing physics.
\par 
  As an example we can have a quick look at String Theory. What is
 String Theory? What is it about? Is it the science of one dimensional hits?
 Do string theoreticians expect to see one day or another spaghetti hits
 in a detector? Is String Theory only a mathematical trick to
 have more degrees of freedom for the maths between the input and
 output states by having, in any case, the goal of calculating
 the probability distributions of zero-dimensional hits?
 If the idea of zero dimensional objects is already ruled out
 by BAB, does it make sense to look for a microphysics based on objects
 of one, or even more, dimensions? What is sure is that if BAB is right,
 it would be highly surprizing that a String Theory for microphysics
 turned out to be right!

\subsection {The right question: what are the apparatuses in GR?}
  Could HAT and the VA help in going farther on this problem of unification?
 We can easily answer "yes", simply by asking the question: what are the
 apparatuses in Special Relativity (SR) and GR? When reading Einstein,
 for example \cite{ref-einstein}, the response to this question is quite
 simple; the apparatuses in SR and GR are\ldots sticks and clocks! SR and GR
 are based on the idea of a space-time continuum that assumes that, whatever
 the geometric scale, we can assign a coordinate quad $(x,y,z,t)$ to all events
 and also to all space-time points. The assignment of coordinates
 done by one observer defines a coordinate frame. A frame is nothing more
 than the piece of maths representing a measurement apparatus in SR and GR.
 Moreover, these theories assume that we can assign two coordinate
 quads to one and the same event, in particular from two frames
 representing two "observers" in motion relative to each other. 
 This kind of double assignment of quads to a same event is at the
 core of the encoding of relativism in SR and GR.
\par
  Having two quads, we pass from one to another with a "transformation".
 In SR, it is the Lorentz transformation (LT) that represents observers in
 a uniform movement relative to each other. In GR, it is a general
 $r^\mu(x0,x1,x2,x3)$ transformation representing any kind of relative
 movement. By using tensors, the SR and GR formalisms make it possible to write
 an expression describing a physical law in such a way that the expression
 stays the same, has same form, after transformation. This constancy of
 form represents, in the maths, the idea of relativism that says that the laws
 of physics should look the same whatever we observe the "out there".
\par
  This is great, but the absolute coordinate assignment
 is\ldots a myth! We cannot build a detector covering all space-time
 for all scales, that is too idealistic. At micro scale, we cannot use a stick
 to do measurements within a presumed "atom" object.
\par
  Moreover, the idea of a double assignment of quads to a same event does not
 hold either at micro scale. Supposing we keep the idea that light is made of
 hypothetical photon objects, most of the time a measurement on one photon,
 for example done with a photomultiplier (PM), is said to destroy the photon.
 So, in such a theory "with photons", we may assign a quad to a photon in
 the frame defined by the PM, but we can no longer associate a quad
 to the same photon from another moving PM since the photon
 object\ldots no longer exists!
\par 
  This loss of double assignment is much more striking with an
 apparatus-centered point of view. In this case an event is a hit
 which is, by definition, attached to a cell of a detector, so a hit cannot
 be attached to two detectors, whether they are in movement or not relative
 to each other. The loss of multiple quad assignments is natural here.
 HAT comes straight in with the right point of view and the consequences
 are drastic.
\par
  Since the "observation" (therefore a coordinate assignment) of a same
 "flash of light" (punctual event) from two different "observers" (frames)
 in movement relative to each other (for example, one observer in a train
 and the other on the platform) is the starting point of Einstein's reasoning
 that leads to SR and then GR, it must be concluded that if, at micro scale,
 we can no longer do this "observation" (quad assignment), then it is the whole
 SR and GR that collapse like houses of cards at this scale
\footnote{with one observer in the Hogwarts Express and the other on the Platform $9\fract_HP3/4$ !}
.
\par
  The loss of double assignment destroys the way that relativism is encoded
 in "frame based theories". For example, there is no reason for the maths
 of micro scale to be "Lorentz covariant" anymore. If the LT makes no sense,
 the idea of "constancy of speed of light" at micro scale is highly
 questionable because the LT was introduced to encode this constancy
 in the formalism. Such questioning about light concurs with the BAB
 argumentation that tells
 us that the word "photon" cannot be associated with a corpuscle "out there":
 how can we speak about the speed of something if there is
 no\ldots "something"!?
\par
 The word "light" should be associated (as should "gravity") to a meso scale
 phenomenon. At this scale, within the Maxwell theory, we can associate
 a speed to this phenomenon which is modeled with waves. With a meso-macro
 scale theory based on
 multiple quad assignments to a single "flash of light" punctual event and the
 axiomatic idea that the "speed of light" is constant for all coordinate
 frames, we can build SR and GR. At micro scale, the best that we can do
 is to associate to this "light" phenomenon the word "photon" defined by
 a pair of $(hit\mbox{-}source,detector)$, but the connection of this
 pair to the
 word "light" of meso-macro scale is now far from being\ldots luminous!
 One idea to achieve this association would be to define the word "lamp"
 as some kind of aggregation of photon-hit-sources. So a "lamp" would be a
 "source of light". The justification of such an aggregation brings us
 to the question of the constitution of apparatuses, a point which is going
 to be discussed later.
\par
 We also start to see how some unifying maths could operate; by keeping
 a frame logic for meso and macro scales (and so keeping SR and GR here),
 but by being able to evolve this maths to a logic not based on frames at
 micro scale.
\par
 At this point an important question arises: if we discard frames, and
 therefore
 SR and GR for micro scale, do we lose completely the idea of relativism at
 this scale?

\subsection {Relativism with HAT and the VA}
 The idea of relativism is that physics laws should be expressed in the same
 manner whatever the way we observe nature. This sounds like a
 great idea and it would be a pity to lose it.
\par
 As seen above, in SR and GR this idea is mapped in the covariance of tensors
 that makes it possible to have a constancy in the form of formulas
 representing physical laws. The idea of relativism in SR and GR is then
 attached to a very peculiar way to "observe". The idea
 of "observation" is attached to the fact of being able to assign
 coordinate quads to everything (then define a "frame" identified
 with the "observer") and to the fact of being able to assign two quads
 to the same event. Looked at from this point of view,
 this is a very particular manner of expressing relativism which comes
 from a too idealistic way of "observing".
\par
 Now if we can no longer keep the frame logic for microphysics, what happens
 to the idea of relativism? Is it possible to define it without frames?
 It appears that we can do so quite easily with an apparatus-centered point
 of view. In HAT, it is sufficient to state:
\begin{quote}
   Whatever the apparatuses layout, the way to calculate the hit
  probability distributions must be the same.
\end{quote}
 This is straightforward, simple. We call the above statement
 the Apparatuses Relativism Principle (ARP). How could it be represented
 in the maths? In fact it appears
 that the HF already does that! Yes, because whatever the apparatuses setup is,
 we attach a $\psi$ to a hit-source apparatus and an operator to a
 detector-apparatus, and we have the same mathematical mechanism to get
 the probability distribution. We have to evolve the $\psi$, then
 decompose the evolved $\psi$ to the operator local $\psi$s
 attached to each possible detector outcome, and then
 take the square modulus of each term of the decomposition to get the
 probability distribution (Born's recipe). We already have a mathematical
 transcription of the idea of relativism in terms of apparatuses and hits for
 microphysics! Moreover, we see that this way of dealing with relativism is
 much more physical than that of SR and GR, because it deals with
 the idea of "observation" in a much more accurate and physical way
 than what is done with sticks and clocks in SR and GR.
\par
 In fact, we may even say that Quantum Mechanics, if understood as a
 hitology, is already much more relativistic than the two relativities
 themselves! (at this point Einstein definitely turns in his grave!)
\par
  It is interesting to note that in various unification attempts, theoreticians
 still keep the whole-coordinate-assignment idea and stay with the
 maths of tensors (extended with spinors) for microphysics. This may make
 sense from a maths point of view but it does not appear to be grounded
 from the point of view of microphysics. If we have to seek for
 new maths it should be for maths that keeps or restores the
 "multiple coordinate assignment of everything" for meso-macro scales, but
 goes to a HAT+VA+ARP+HF logic for micro scales.

\subsection {A key experiment related to gravity in microphysics}
  To make progress on gravity at micro scale knowing that there are, because
 of BAB, huge questionings about the idea of objects at this scale, we have to
 do experiments that pose the right questions. Probably the best one
 that we can imagine would be a Bohm-Aspect setup with a gravity component,
 for example by introducing a "massive object" close to the
 "line of flight" (!) between the hit-source and one of the detectors.
 So an experimental setup with the three cases
\begin{enumerate}\addtolength{\itemsep}{-0.5\baselineskip}
   \item no massive object on either arm.
   \item one massive object close to one arm.
   \item one massive object close to each arm.
\end{enumerate}
 If we label the three setups with $\sigma$=1,2,3, then we could get
 runs $C_{exp}(\theta,\sigma=1,2,3)$. What would the experimental
 curves be? Do we have a theory that could model this to
 give a $C_{some\ theory}(\theta,\sigma)$ to compare with
 $C_{exp}(\theta,\sigma)$?
 Moreover, an ideal situation would be massive objects able to induce a gravity
 effect interpretable with GR, then interpretable as a space-time effect.
 For example, some\ldots black holes would be nice!
 (Micro-macro experimental physics at last!).
 Here we would truly mix gravity with questions about microphysics.
 We would learn a lot about gravity in microphysics here!
\par
 This kind of Bohm-Aspect-Einstein setup would be the most
 interesting since it would be related to the Bell reasoning.
 Something similar done with a variant of the two-slits or MachZender setups
 by putting, or not, massive objects close to their "lines of flight" would
 be very interesting too.

\section {What are apparatuses made of? The true fundamental question}
\subsection {QM textbooks?}
  With QM textbooks, we cannot answer this question because the word
 "particle" is not mapped to a direct maths symbol as a trajectory, and
 therefore we have nothing to recover a $XYZ_{body}(t)$ describing a body at
 our scale: we cannot build something from\ldots nothing! 
\par
  In QM textbooks, this question is related to subjects as the
 "measurement problem", the "quantum to classical transistion" and
 the "decoherence". A lot is written about them, but we don't quite
 see how physicists can make progress without some reliable micro entity
 to build on!
\par
  For example, for us "decoherence" is, first of all, a mathematical
 manipulation within the HF that shows that a "density matrix" (a version
 of $\psi$) can evolve to become diagonal. Fine, so what? If the density
 matrix is still not mapped to an ontological entity we have made no progress!

\subsection {Bohm-model, Consistent Histories?}
 We may look for other interpretations that explicitly restore "particle"
 as an axiomatic idea, because with the idea of particle, and equipped with an
 aggregation mechanism, it is possible to recover bodies (a $XYZ(t)$ in
 the maths). However, we have seen already that attempting to keep particles
 along with the HF leads in general to the introduction of additional
 weird ideas. In the Bohm model, in which there is a $XYZ(t)$, the weirdness is due to action at a distance.
\par
 Another candidate model is the "Consistent Histories" (CH in short).
 At least in \cite{ref-gell-mann} and \cite{ref-omnes}, CH is presented
 with "particle" as an axiomatic idea. The compatibility with the HF is
 restored at the cost of axiomatic constraints in the method of calculating
 probabilities over possible "histories". In CH, the pruning of "branches"
 makes it possible to recover physical bodies and "us", as human beings\footnote{In CH, we are IGUSes! IGUS for Information Gathering and Utilizing System. To this respect, "We are OO thinkers!" looks, for us, a more relevant statement to discuss with.}.
 This seems appealing, but looking closer it is still weird 
 and the weirdness is related to these axiomatic constraints.
 From what we understand (?), there are objects that can have a
 property (position) in a first set of histories, but may not have
 this property (i.e. no position but momentum) in another set of histories
 said to be "not consistent" with the first set. The constraints allow
 avoiding the assignment of a join probability to sets that are not compatible
 (in particular not to assign a probability to "position and momentum").
 So the constraints are related to the fact that "out there"
 there are objects for which we can no longer say if they have a
 property or not! If this is the case, then it is here that our intuition
 rebels because this is definitely\ldots anti OO! In OO, an object has 
 a property or it does not. If we look for an OO interpretation of the HF,
 CH is not the right horse to back.

\subsection {HAT?}
  In HAT the situation is de facto clean. The question 
 "what is a hit-source or detector apparatus made of?" is simply a bad
 question in HAT. It is a bad question because apparatuses belong
 to the axiomatic ideas and as such they cannot be built
 from more basic ideas. We may think that apparatuses can be built
 back from the VA, but this is not possible since the VA is deduced from
 apparatuses. So within HAT+VA the question is ill defined.
 We claim that being able to recognize this is progress
 compared to QM textbooks because QM textbooks cannot clearly state
 whether the question is ill defined or not within their axiomatic ideas.
 Logically a question can have an answer, but it may also be ill defined
 in such a way that no answer is possible, and being able to see that
 a question is ill defined is progress.
\par
  Nevertheless, we now come across another problem which is our HAT rereading
 of the "measurement problem" in QM. Agreed, by recovering understandable ideas
 at all stages in HAT, we now have a very consistent approach to deal with
 microphysics, but this situation is not satisfactory (which is not, for us,
 the same as uncomfortable) because all our intuition tells us that
 an apparatus is made of something. 
 We can take a hammer and smash an apparatus to pieces. What is
 then the status of the pieces? At the very moment when they cease to be
 a detector, do they belong to the VA? Are they new hit-source apparatuses
 that "do nothing"? Despite the fact that a theory based
 on the axiomatic ideas of the quad $(detector,hit,hit\mbox{-}source,\mbox{VA})$
 is highly consistent it is still frustrating because
 it is not intuitively satisfactory at the level of the status
 of the apparatuses. All our intuition tells us that the apparatuses
 cannot be axiomatic ideas. Which leads us
 to our rephrasing of the "measurement problem" in HAT that we state as
\begin{quote}
   The "measurement problem" is the dissatisfaction
  that today's microphysics leads us toward a consistent model (HAT) in which
  apparatuses are axiomatic ideas, whilst our intuition tells us that
  they are not!
\end{quote}
  We see that this problem can be rephrased in a more OO and
 therefore more comprensible way that allows us to ask the right questions.
 In particular, a straightforward one is: can we build more OO axiomatic
 ideas so that these new axiomatic objects can recover the
 $(detector,hit,hit\mbox{-}source,vacuum\mbox{-}of\mbox{-}apparatuses$)
 of our hitology?

\subsection {"aparticles"?}
  Noticing that the BAB argumentation (if granted) rules out corpuscles only
 for the "in between" apparatuses, one way to build more OO axiomatic ideas 
 would be to introduce some
 corpuscles or particles as an axiomatic idea but dedicated
 to build "apparatuses only". If we name this kind of corpuscle
 "aparticle" (for apparatus particle, or aggregatable particle),
 the axiomatic ideas would be: aparticle and the VA. Then
 the hit-source and detector apparatuses would be made by an aggregation of
 aparticles. We could perhaps even restore an $(emitter,propagation,impact)$
 logic by saying that inside a hit-source, an aparticle
 "does something" to the VA, that this VA modification is seen as a
 propagation at our scale up to the detector in
 which an aparticle reacts producing a hit, with the maths describing
 what happens between the emission and the impact being the HF.
\par
  Here's an idea; what if our brand new aparticle were nothing more
 that the\ldots "good old atom"?! If so, this "atom" should be equipped
 with some special property so that it can be seen, from a HAT point of view,
 as an axiomatic idea to build back an apparatus.

\subsection {Pairs of $(emitter,receiver)$?}
  Let us consider another idea. We have also seen that a "particle" could
 be defined by a couple $(hit\mbox{-}source,detector)$ in HAT. A hit-source
 without a corresponding detector is nothing, and a detector without
 its hit-source is nothing either. So would
 it be possible to push the idea further and say that the "out there"
 is constituted by elementary pairs of $(hit\mbox{-}source,detector)$ or
 $(emitter,receiver)$ and that these are the fundamental
 building blocks of everything? Assuming that a gathering
 of micro pairs is possible, we would be able to recover a
 meso $(hit\mbox{-}source,detector)$ pair, but would also 
 recover the idea of being able to smash it into pieces.
\par
 To build a consistent model of elementary pairs, we would have
 to find some dynamics for these elementary pairs; at least some kind
 of crystalization dynamics to recover our scale $(hit\mbox{-}source,detector)$
 and some kind of "soup dynamics" to recover the VA from special
 states of micro pairs.

\section {Relationship with software}
  OO is a terminology coming from software and it is not a coincidence
 if it is used here. We can have a close view of how data is treated
 if working a little on software for HEP experiments. This makes us aware
 that the primary input of "all that" is nothing more than a bunch
 of hits appearing in detectors if the right conditions are met,
 and in particular if we have, at last, built the right 27 km long
 accelerator! Software in HEP makes us realize that trajectories and particles
 are secondary entities that are "recovered back" in a step of data
 treatment called "reconstruction", a step which is targeted
 to build back the "final state", the one after the "interaction" (dangerous
 vocabulary). Here we see that particles and trajectories are clearly something
 introduced by us, humans, when treating data.
\par
  To treat data, we need
 to write software and to do that we need some programming language.
 For a long time HEP computer programs were done by using procedural
 languages such as FORTRAN, but the 1990s saw the migration to object
 oriented (OO) programming. This kind of language puts a gun in one's back and
 compels us to think hard to know "what are the classes", what could
 be considered as objects defined as standalone entities bearing their own
 properties. For the author, at a certain moment this questioning applied
 to HEP data treatment ran into the questioning about his longstanding
 discomfort with QM. This questioning leads to the conviction that one
 key ingredient to understand microphysics (the missing link!) is the
 recognition that our way of thinking is naturally strongly OO, that
 we are "OO thinkers". From here
 it is quite easy to reach the conclusion that perhaps the number one issue
 with the QM interpretation problem is that we have missed some key objects
 or key classes in the whole story.

\section {Conclusions}
\subsection {Summary of the overall argumentation}
  After having defined what "doing physics" meant for us, we decided
 to examine the situation in microphysics.
  By granting the Bohm-Aspect experiment results and the Bell reasoning
 (BAB), we acknowledged that the concept of "corpuscle flying around" is hardly
 tenable for microphysics. We have seen that Quantum Mechanics (QM),
 defined as the Hilbert Formalism (HF) attached to a particle vocabulary
 is intuitively misleading. We have seen that
 reinterpreting the HF in terms of detector apparatuses, hits in detectors
 and hit-source apparatuses is a much more natural
 interpretation, especially if we have in mind how data is treated
 in experiments. We have named this interpretation the HAT interpretation.
 This is a more natural
 interpretation because it is object oriented (OO) and
 OO is a natural driving paradigm for us. By using the three
 OO concepts: detector apparatus, hit, hit-source apparatus, 
 and a reinterpretation of "decay" experiments
 we have revealed the existence of the in-between apparatuses entity
 or vacuum of apparatuses (VA) entity, defining it in a very clear
 way for us humans. We reached the conclusion that the VA is an active
 entity having the capability to transmute hit patterns. 
\par
 We granted that the maths of Quantum Field
 Theory (QFT) is the right one to describe the VA, but we rejected
 any particle vocabulary attached to a QFT.
 Because of hit pattern transmutation, we saw that the VA cannot
 be reduced to a space-time geometry, inducing
 that all theories, as General Relativity (GR), which are
 based only on a space-time geometry for the "in between" entity cannot be good
 for microphysics.
\par
  We saw that the problem of unification
 of Quantum Mechanics and General Relativity, if taken only at the
 level of the maths, is doomed, and that it has, first, to be solved
 at the level of the ideas. We saw that SR and GR are based
 on a too idealistic conception of a measuring apparatus. We saw that
 the reconsideration of the idea of apparatus at all scales could lead
 to the right underpinning maths able to recover the HF at
 micro scale and a Riemann geometry for meso and macro scales.
 Along the way, we stated an apparatus based relativism principle (the ARP).
\par
  We ended by rephrasing the problem
 of measurement in QM as the question "what is an apparatus made of?"
 and saw that it is an ill defined question in HAT. To overcome
 the frustration of not being able to "smash apparatuses", we
 mentioned two ways to build models with axiomatic ideas that can
 recover those of HAT, the first based on the aparticle idea and
 the second based on the idea of elementary pairs of
 micro $(emitter,receiver)$.

\subsection {Is nature OO?}
  So, to the question "Is nature OO?", we can answer that for meso and macro
 scales the answer is "yes". For microphysics, it looks like we have to yield
 some ground, but we believe that the answer is not yet "no"; a hitological
 point of view is a card still to be played. What is sure is that,
 if nature is truly not OO at this scale, "doing physics" will become
 a weird $(maths,experiments)$ couple activity, with no clear meaning for
 us because no longer grounded on\ldots reliable ideas!

\subsection {Yes, we can!}
  About the famous quote of R.Feynman \cite{ref-feynman}:
\begin{quote}
  I think I can safely say that nobody understands Quantum Mechanics
\end{quote}
 we say: if Quantum Mechanics is understood as the
 Hilbert Formalism attached to a particle vocabulary then yes, we agree,
 we really don't see how someone can have a full
 understanding of Quantum Mechanics. Now to the question:
\begin{quote}
  Can we understand microphysics?
\end{quote}
 we say: yes, we can! By using a hitology and a "vacuum of apparatuses"
 we can restore an Object Oriented point of view which, associated to the
 Hilbert Formalism, makes it possible to still understand this kind of physics.

\subsection {No maths, only ideas}
  Some may note that there is no maths of our own in this text. This was
 done deliberately because of the deep conviction
 that the number one problem in today's physics is more around
 the ideas than around the inflationary maths. We hope that the chain of
 reasoning and ideas found in this text may help those who
 have intuitive discomfort with microphysics. What is sure
 is that the author, with his new HAT, sleeps much better now!

\section {Thanks}
  Thanks to Marcel Urban and the LAL MEDOC group for including me in
 some of their discussions. This led, in October 2009, to a first
 formulation, for a slide presentation in French,
 of thoughts which had been brewing for a long, very long time.
 This crystalization material led to this text, in English, in July 2011.
\par
  Thanks to Michel Bitbol; his book \cite{ref-bitbol} helped me a lot
 to realize that my "longstanding intuitive discomfort" was not a
 problem requiring\ldots psychoanalysis!
\par
  Special thanks to Jane Moneypenny for her help with the language
 of\ldots Newton!

\end{document}